\documentclass[twocolumn]{aastex61}
\usepackage{amsmath}

\defcitealias{2017ApJS..230....4F}{F17}
\newcommand{\msun}{\ensuremath{\mathrm{M}_{\odot}}}
\newcommand{\ptfR}{\ensuremath{R_\mathrm{P48}}}

\shorttitle{The volumetric rate of calcium-rich transients}
\shortauthors{Frohmaier et al.}

\begin{document}

\title{The volumetric rate of calcium-rich transients in the local universe}

\email{chris.frohmaier@port.ac.uk, m.sullivan@soton.ac.uk}

\author[0000-0001-9553-4723]{Chris Frohmaier}
\affiliation{Department of Physics and Astronomy, University of Southampton, Highfield, Southampton, SO17 1BJ, UK}
\affiliation{Institute of Cosmology and Gravitation, University of Portsmouth, Portsmouth, PO1 3FX, UK}

\author[0000-0001-9053-4820]{Mark Sullivan}
\affiliation{Department of Physics and Astronomy, University of Southampton, Highfield, Southampton, SO17 1BJ, UK}

\author[0000-0002-9770-3508]{Kate Maguire}
\affiliation{School of Mathematics and Physics, Queen's University Belfast, Belfast BT7 1NN, UK}

\author[0000-0002-3389-0586]{Peter Nugent}
\affiliation{Department of Astronomy, University of California, Berkeley, CA, 94720-3411, USA}
\affiliation{Lawrence Berkeley National Laboratory, Berkeley, CA, 94720, USA}
\begin{abstract}

We present a measurement of the volumetric rate of `calcium-rich' optical transients in the local universe, using a sample of three events from the Palomar Transient Factory (PTF). This measurement builds on a detailed study of the PTF transient detection efficiencies, and uses a Monte Carlo simulation of the PTF survey. We measure the volumetric rate of calcium-rich transients to be higher than previous estimates: $1.21^{+1.13}_{-0.39}\times10^{-5}$\,events\,yr$^{-1}$\,Mpc$^{-3}$. This is equivalent to 33--94\% of the local volumetric type Ia supernova rate. This calcium-rich transient rate is sufficient to reproduce the observed calcium abundances in galaxy clusters, assuming an asymptotic calcium yield per calcium-rich event of $\sim$0.05\,\msun. We also study the PTF detection efficiency of these transients as a function of position within their candidate host galaxies. We confirm as a real physical effect previous results that suggest calcium-rich transients prefer large physical offsets from their host galaxies.

\end{abstract}


\section{Introduction}
\label{sec:intro}

Over the last decade, all-sky optical time-domain surveys have uncovered a new class of so-called `Ca-rich' transients \citep{2010Natur.465..322P,2012ApJ...755..161K,2014MNRAS.437.1519V,2017ApJ...836...60L}. These events have properties intermediate to those of supernovae (SNe) and novae, with characteristics that set them apart from the classical SN population: absolute magnitudes in the range of $-15$ to $-16.5$\,mag, a rapid photometric evolution with rise times of 10--15 days, photospheric velocities of 6000--11000\,km\,s$^{-1}$, and a fast spectroscopic evolution to calcium-dominated nebular spectra.

The number of published events remains small: \citet{2017ApJ...836...60L} considered a combined sample of 8 events with good photometric coverage. These objects all occurred in the outskirts of early-type galaxies, at $\sim$8--80\,kpc from the centers, and a tentative preference for cluster environments. This suggests a metal-poor (and perhaps also old) explosion environment, ruling out a massive-star origin if they are formed in situ. \cite{2013MNRAS.432.1680Y} showed that their locations are consistent with globular cluster distributions. However, photometric searches for globular clusters at the positions of known Ca-rich transients have, on the whole, been unsuccessful \citep{2014MNRAS.444.2157L,2015MNRAS.452.2463F,2016MNRAS.458.1768L,2017ApJ...836...60L}. 

Several progenitor scenarios have been suggested to explain the physical origin of Ca-rich events. These include helium-shell detonations (without core detonations) on the surface of low-mass carbon-oxygen (CO) white dwarfs \citep{2011ApJ...738...21W}, collisions in binary systems involving a He donor and a CO or ONe white dwarf \citep{2017MNRAS.468.4815G}, and suggestions that they may result from the tidal disruption of CO white dwarfs by intermediate-mass black holes \citep{2009ApJ...705L.128R,2012MNRAS.419..827M,2015MNRAS.450.4198S}.

Ca-rich transients may also play an important role in other areas of astrophysics. A long-standing puzzle is the origin of the observed Ca/Fe over-abundance in the intra-cluster medium \citep[ICM;][]{2007A&A...465..345D,2016A&A...592A.157M}, which cannot be explained by the yields of traditional SN classes such as core-collapse SNe and SNe Ia. \cite{2014ApJ...780L..34M} showed that a contribution from Ca-rich transients could provide a source of calcium to match the ICM measurements. The remote locations would also make them efficient at polluting the ICM, since their ejecta does not have to escape a galaxy potential \citep{2004ApJ...613L..93Z}. The source of Galactic positrons has also been suggested to be due to faint thermonuclear SNe that produce the positron emitter $^{44}$Ti, with the decay of $^{44}$Ti leading to the production of $^{44}$Ca, either through the merger of two white dwarfs \citep{2017NatAs...1E.135C}, or as a natural by-product of He-shell detonation \citep{2011ApJ...738...21W}. 

However, the rate of Ca-rich transients is uncertain due to the small numbers discovered to date. This is a byproduct of the faintness and speed of their light curves which makes them difficult to detect. Absolute volumetric rates have not been calculated for a homogeneous sample of Ca-rich transients before; rather, rates relative to type Ia supernovae (SNe Ia) have been inferred. \citet{2010Natur.465..322P} inspected the spectra of objects in the Lick Observatory Supernova Search (LOSS) \citep{2001ASPC..246..121F} for possible Ca-rich candidates. An incompleteness assumption was made to produce a sample of 2.3 Ca-rich transients and 31 SNe Ia in the same search. They concluded the Ca-rich rate is some 7\% of the SN Ia rate. Similarly, \citep{2012ApJ...755..161K} placed a lower-limit on the Ca-rich rate by comparing the number of Ca-rich events discovered in the Palomar Transient Factory \citep[PTF;][]{2009PASP..121.1334R} to the number of SNe Ia in the same volume. Without any assumption of PTF's incompleteness, they estimated that the Ca-rich transient rate must be ${>}$2.3\% of the SN Ia rate. A proper treatment of the survey incompleteness will also help explain the apparent preference for remote locations due to a potential for selection biases in their discovery (for example, it is  easier to discover transients on fainter backgrounds away from galaxies).

The aim of this paper is to determine the absolute volumetric rate of Ca-rich transients, using a sample from PTF, and to determine if they have a preference for the outskirts of galaxies. PTF was an automated optical sky survey operating at the Samuel Oschin 48 inch telescope (P48) at the Palomar Observatory, specifically designed for transient detection. This paper builds on a comprehensive study of the PTF detection efficiencies presented in \citet[][hereafter \citetalias{2017ApJS..230....4F}]{2017ApJS..230....4F}, and applies them to a controlled sample of PTF Ca-rich events to estimate the first absolute rate of these events. Throughout, we assume a flat $\Lambda$CDM cosmological model with $\Omega_\mathrm{M}=0.3$ and a Hubble constant of $H_0=70$\,km\,s$^{-1}$\,Mpc$^{-1}$.

\section{The Ca-rich sample}
\label{sec:sample}

A parent sample of Ca-rich SNe in PTF was presented by \citet{2017ApJ...836...60L}. This identified four likely members of the class  \citep[PTF11kmb and PTF12bho, along with PTF10iuv and PTF11bij previously identified by][]{2012ApJ...755..161K}. Their sample also included PTF09dav \citep{2011ApJ...732..118S}. However, we have excluded this object as it displays unusual photospheric spectra, and late-time spectra that include hydrogen emission \citep{2012ApJ...755..161K}, and thus is likely not a member of the same class. 

\citetalias{2017ApJS..230....4F}, and thus our study, is further restricted to the PTF survey period 2010--2012 (note this would also exclude PTF09dav), and to periods when PTF was observing in the $R$-band filter (\ptfR); around 85\% of the survey. We also require all objects in the sample to have been observed on at least four nights from $-$15 to $+$30 days from peak brightness conventionally defined as day zero. PTF11bij was observed mainly in the PTF $g$-band filter and has insufficient \ptfR\ data to pass this requirement, and is thus excluded from our analysis. Our final sample therefore has three events.

\begin{figure*}
\centering
\includegraphics[width=1.0\linewidth]{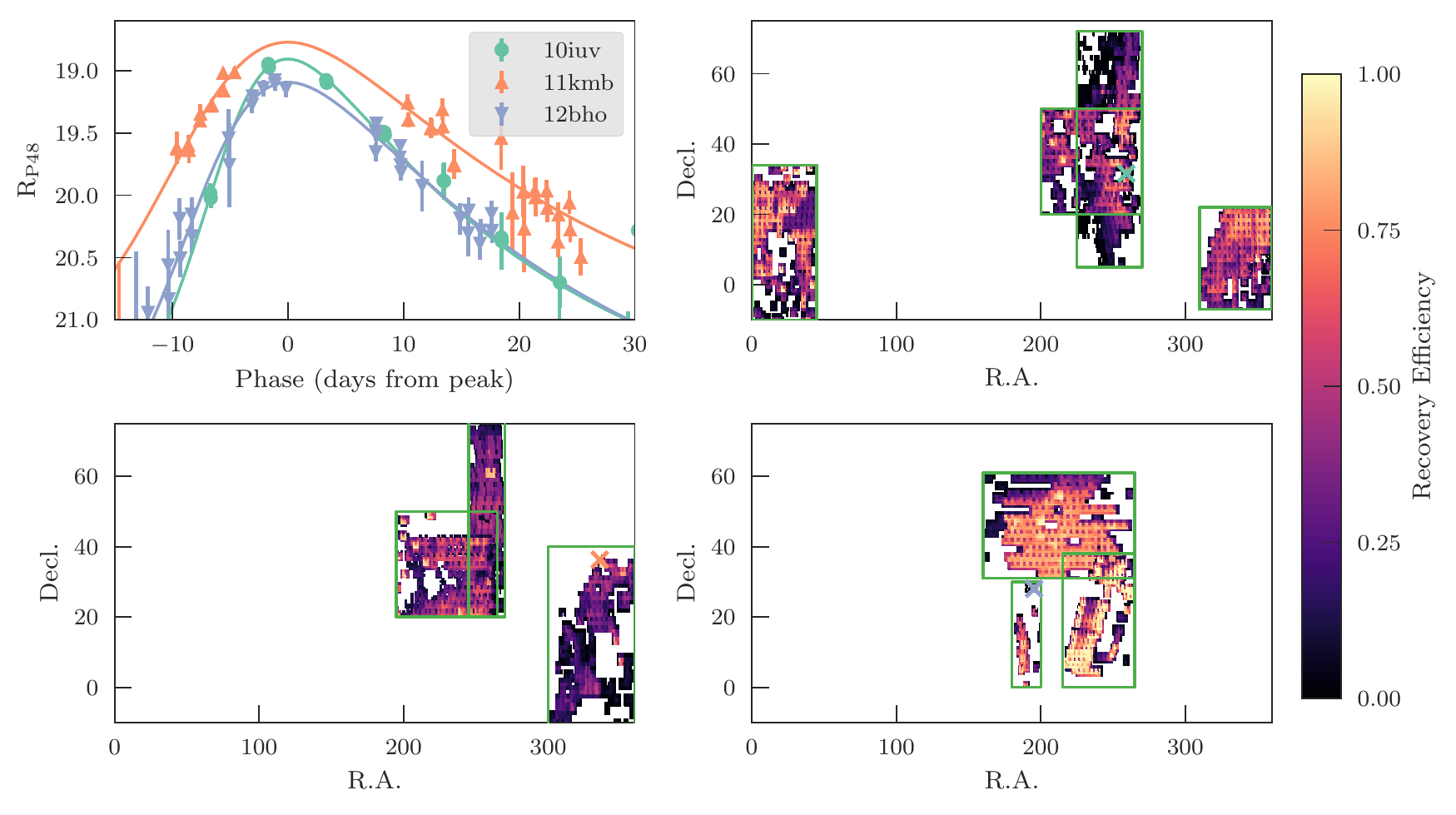}
\caption{The three calcium-rich transients entering our sample, and their positions on the sky. Top left: The three light curves in apparent magnitude space, together with the \citet{2009A&A...499..653B} functional fits over-plotted. The other three panels show the PTF survey area considered in each year: 2010 (upper right), 2011 (lower left), and 2012 (lower right). In each panel, the simulated areas are shown as green boxes, and the position of the calcium-rich transients as crosses. The color-coding shows the fraction of simulated events that passed the coverage cuts.}
\label{fig:ca_lc_sky}
\end{figure*}

The light curves of these three events are shown in Figure~\ref{fig:ca_lc_sky}, along with smoothed \ptfR\ light-curve templates fitted using the phenomenological functional form of \citet{2009A&A...499..653B}, which is sufficient to fit the shape of most smoothly-evolving supernovae. These model fits will be used in our simulations to model the Ca-rich events.

To determine the PTF spectroscopic completeness of Ca-rich transients, \citet{2017ApJ...836...60L} searched the PTF database for all objects that had luminosities between those of SNe and novae. They found one additional transient that may be a Ca-rich event (PTF10hcw), but its sparse light curve means it does not pass our light curve quality criteria. We note that if further objects were unconfirmed in the PTF database, this would have the effect of increasing the rate of Ca-rich transients that would be determined.

\section{The Rate of Ca-rich SNe}
\label{sec:rate}

We next outline our method for determining the Ca-rich transient volumetric rate. The challenges are our small sample size, the large PTF observing footprint, and the variable cadence of the PTF observing strategy. We experimented with an `efficiency-based' method, weighting each event by its overall detection efficiency, as often used in high-redshift SN rate calculations from well-controlled surveys \citep[e.g.,][]{2012AJ....144...59P}. However, the small number of events in our sample, and the highly-variable and sky position-dependent detection efficiencies of our events, made this approach unreliable.

Instead, we used a Monte Carlo technique similar in concept to that of \citet{2017MNRAS.464.3568P}. Given an input intrinsic volumetric rate of events ($r_\mathrm{input}$), we simulate how many events PTF would have expected to detect based on its observing strategy ($N_\mathrm{obj}$). By varying this input volumetric rate over a wide range of values, and determining how often $N_\mathrm{obj}=3$ in the simulations, we can construct the probability distribution of the volumetric Ca-rich rate, $r_V$.

Key to this approach is the PTF detection efficiency study of \citetalias{2017ApJS..230....4F}. This study simulated around 7 million point sources in the PTF survey, and studied their detection efficiency by the PTF pipeline as a function of source magnitude, host galaxy surface brightness, and various observing conditions (the seeing or image quality, the sky brightness, and limiting magnitude). From this, multi-dimensional detection-efficiency grids were constructed, allowing the detection efficiency of any point source in any PTF image (and on any arbitrary host background) to be calculated.

Our detailed procedure is as follows. We first set the area and time period over which PTF would have been sensitive to Ca-rich events. We chose 10 observational footprints from PTF over 2010--2012 with a total of 9,428\,deg$^2$ of sky. Each field was observed in a rolling cadence by PTF for between 30 and 155 days ($\mathrm{T}_i^\mathrm{obs}$; $i$ denotes the field, from 1 to 10). These fields each have a volume $V_i$ over the redshift range $0.0035\le z \le0.03$, inside which PTF was sensitive to calcium-rich events. This volume is determined at the lower end by the redshift at which the calcium-rich transients would saturate the PTF detector ($z=0.0035$), and at the upper end by the redshift at which the faintest of the calcium-rich transients would no longer be detected by PTF ($z=0.03$).

A value for $r_\mathrm{input}$ is randomly chosen from the interval $1{\times}10^{-6} \le r_\mathrm{input} \le 5{\times}10^{-5}\ \mathrm{events}\ \mathrm{yr}^{-1}\ \mathrm{Mpc}^{-3}$, and used to randomly draw a number of events, $N_\mathrm{input}$, from a Poisson distribution
\begin{equation}
P(N_\mathrm{input}; \lambda)=\frac{\lambda^{N_\mathrm{input}} e^{-\lambda}}{N_\mathrm{input}!}
\end{equation}
where $\lambda=r_\mathrm{input}\sum\mathrm{T}_i^\mathrm{obs}V_i$ and the sum runs over the 10 fields. Each of these $N_\mathrm{input}$ events are then randomly assigned a sky position, date of peak brightness, and redshift, such that they were uniformly distributed through the simulated volume and survey duration. Each event was then randomly assigned a Ca-rich light-curve template from Section~\ref{sec:sample}, which was then scaled to the assigned redshift.

Simulations were then performed, replicating the operation of PTF and matching the nightly observing conditions and cadence patterns. Each time PTF `observed' an artificial Ca-rich transient, the \citetalias{2017ApJS..230....4F} efficiencies were used to statistically assess whether PTF would have detected the event at that epoch. The recovered light curves of each simulated event were then checked to determine whether they met the light-curve quality criteria for the real sample: at least four detections over at least four nights. The number of `observed' Ca-rich events, $N_\mathrm{obj}$, that met the criteria, and the associated value of $r_\mathrm{input}$ for that realization, were recorded.

This process was repeated for ${>}1.7{\times}10^6$ realizations of the Ca-rich rate, at which point the statistical improvement from further simulations became negligible. This resulted in the analysis of ${>}3.5\times10^8$ simulated Ca-rich light curves. Clearly, this method of rate calculation is computationally expensive, but the results of the simulation are easy to understand and the uncertainties can be intuitively handled.

\begin{figure}
\centering
\includegraphics[width=1.0\linewidth]{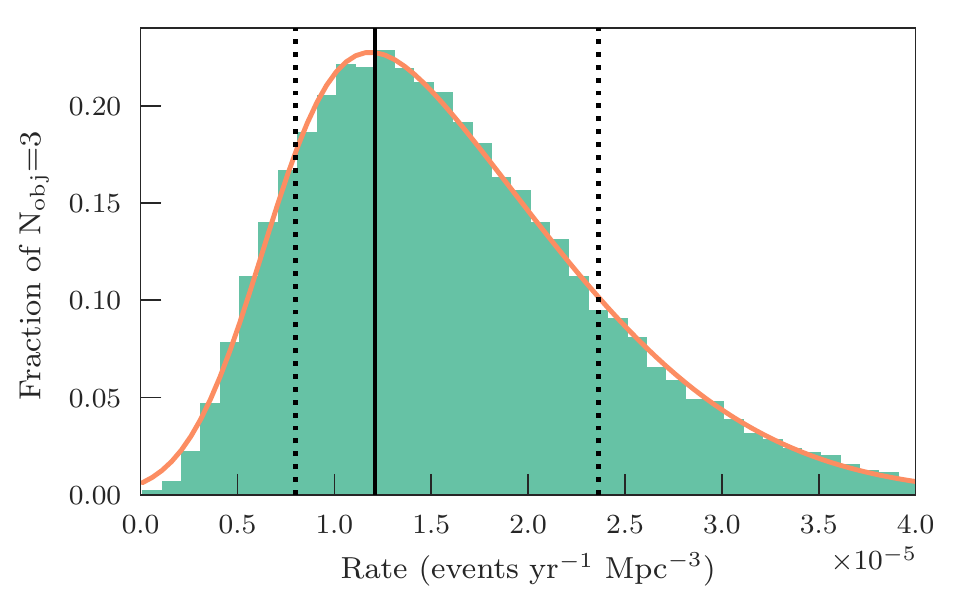}
\caption{The fraction of our simulations that result in three Ca-rich transients being detected by PTF ($N_\mathrm{obj}=3$) as a function of the input volumetric rate $r_\mathrm{input}$. The over-plotted function is a skewed Gaussian, with the solid vertical line denoting the median of the function, and the vertical dotted lines the 68.3\% confidence interval. See Section~\ref{sec:rate} for details.}
\label{fig:ca_rate_dist}
\end{figure}

Figure~\ref{fig:ca_rate_dist} shows the fraction of our simulations that resulted in $N_\mathrm{obj}=3$ as a function of $r_\mathrm{input}$. The distribution is well described by a skewed-Gaussian, and we use this functional form to estimate the most likely value of the Ca-rich rate and the region containing 68.3\% of the probability. The volumetric rate of Ca-rich transients is then
\begin{equation*}
r_V=1.21^{+1.13}_{-0.39}\times10^{-5}\ \mathrm{events}\ \mathrm{yr}^{-1}\ \mathrm{Mpc}^{-3}\ h_{70}^{3},
\end{equation*}

The volume-weighted mean redshift of the simulation was $z=0.023$.

This represents the first direct measurement of the Ca-rich transient rate, and is equal to around $\sim50$\% of the SN Ia rate at a similar redshift (Frohmaier et al., in prep). Combining our rate with assumptions on the performance of the Zwicky Transient Facility (ZTF) \citep{2014htu..conf...27B,2017arXiv170801584L}, we estimate that ZTF should discover more than 20 Ca-rich transients each year. This order of magnitude improvement on the sample statistics will revolutionize the understanding on the volumetric rates, progenitors and environments of these events.

\section{The remote locations of Calcium-rich events}
\label{sec:Eff_Environment}

\begin{figure*}
\centering
\includegraphics[width=0.85\linewidth]{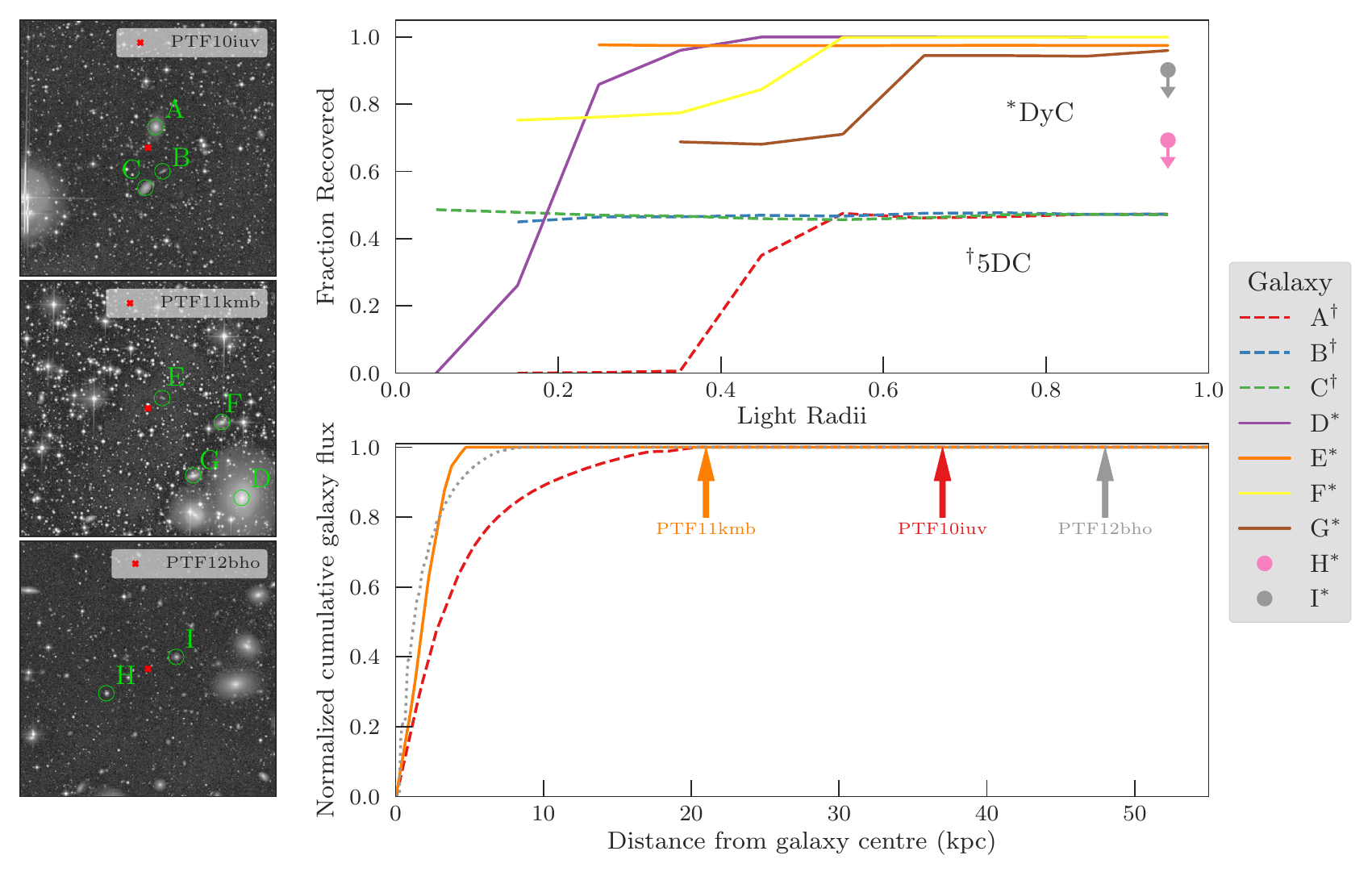}
\caption{Study of the detection efficiency of the three PTF calcium-rich transients in our sample in relation to their host galaxies. Suspected host galaxies are identified in green, and their surface brightness profiles were measured. Ca-rich events were simulated in annuli containing fractions of galaxy light. In the top panel we show that if Ca-rich events followed the light profile then PTF would have recovered more events than it missed. The efficiencies hit an apparent ceiling at larger radii as a result of the different PTF cadence experiments called the `5-day-cadence' (5DC) or the `dynamic cadence' experiment (DyC). The median cadences of PTF10iuv, PTF11kmb and PTF12bho discovery fields were 4.95, 0.96 and 1.00 days respectively. This result demonstrates that our Ca-rich sample is not significantly bias to remote locations, and these transient preferentially occur at large radii. The bottom panel shows the normalized cumulative fraction of galaxy light, as a function of distance from the galaxy core, for the  nominal hosts. The arrows show the distance of each Ca-rich transient from the host, and highlights the remote location of the explosion environment.}
\label{fig:gal_effs}
\end{figure*}

We now consider the efficiencies of recovering a Ca-rich transient as a function of distance from its host galaxy.

We use the real observations of each detected Ca-rich event's location, and the survey performance around the object discovery date, to assess whether each real Ca-rich event would have been discovered had it occurred in a different (but nearby) host environment. This allows at least a partial disentanglement of the intrinsic host environmental effects (e.g., bright galaxies) from observational effects (e.g., cadence and depth), in that we are restricted to survey periods and areas where the data were known to be good enough to discover Ca-rich events.

We begin by measuring the surface brightnesses for the Ca-rich transient host galaxies using high-quality reference images of the fields. The reference images were made from \ptfR\ observations, uncontaminated by SN light. Each image in the stack was resampled to a common coordinate system, and co-added to produce a deep image. Potential host galaxies for our Ca-rich events were identified in the literature \citep{2012ApJ...755..161K,2017ApJ...836...60L} and form the sample for our study. We then measure these galaxies in the images using elliptical apertures, and construct annuli containing fractions of the integrated galaxy light out to a distance of at least 45kpc.

We then simulate each of the three Ca-rich template light curves from Figure~\ref{fig:ca_lc_sky} at a random position inside each of the elliptical annuli for each galaxy. The date of peak brightness for each of the real Ca-rich transients loosely sets the date of our simulated events in the field: for each simulated event, we chose a random date of peak-brightness to be within $\pm20$ days of the real event's peak. We record the local surface brightness and semi-major axis distance to the center of the galaxy for each simulated event. We use the infrastructure of \citetalias{2017ApJS..230....4F} to determine whether each simulated event would have been detected, again applying the requirement of four or more detections.

The results of our simulations -- the fraction of events successfully recovered in each annulus -- are shown in Figure~\ref{fig:gal_effs}. We find that galaxies with the brightest cores show a reduced recovery efficiency. However, under the assumption that the \ptfR\--light approximately traces the galaxy stellar mass, simulated Ca-rich events contained within the outer $\sim50\%$ of the stellar mass are recovered with a near constant efficiency. 

Furthermore, all the Ca-rich events in our sample were found at larger distances than an isophote containing 100\% of the \textit{observable} stellar light (i.e., where the galaxy flux becomes indistinguishable from the background).

The final two galaxies in Figure~\ref{fig:gal_effs} (H and I) had bright cores, such that most of the luminosity was contained in an area smaller than our spatial resolution. We therefore were unable to reliably split our simulations into different elliptical annuli, and instead show the fraction of recovered events within a single boundary containing 90\% of the luminosity/stellar mass.

These simulations demonstrate that if the radial distribution of Ca-rich transients follows the stellar mass of galaxies, PTF would have been capable of finding more events than it missed i.e. the recovery fraction for the speculated hosts is ${\sim}50\%$ or more in an annulus containing 50\% of the stellar mass. Furthermore, we performed a simple Monte-Carlo simulation of Ca-rich events under the assumption they follow the host stellar light profile. We found that ${<}8\times10^{-5}\%$ of the simulations would produce all 3 events beyond the contours enclosing 99\% of the host light. This allows us to confidently reject any hypothesis that suggests Ca-rich transients follow the stellar light. We conclude that the PTF Ca-rich transient sample is not significantly biased due to the performance of the PTF experiment, and that the large separation between event and host-galaxy is an astrophysically real phenomenon. This confirms previous studies that show Ca-rich transients do not follow a host stellar mass profile \citep{2013MNRAS.432.1680Y}, but with a robust consideration of the efficiency of the PTF experiment.

A final consideration on the detection efficiency relates back to the rates calculated in Section~\ref{sec:rate}. If indeed there are Ca-rich transients in the cores of galaxies, where our results suggest PTF has difficulty finding such events, then these objects would increase the absolute volumetric rate we find. Since we derive our Ca-rich rate from a sample of observed events, stricly speaking it should be considered as a lower limit if a significant population of Ca-rich events are later found to occur in galaxy cores.

\begin{figure*}
\centering
\includegraphics[width=0.8\linewidth]{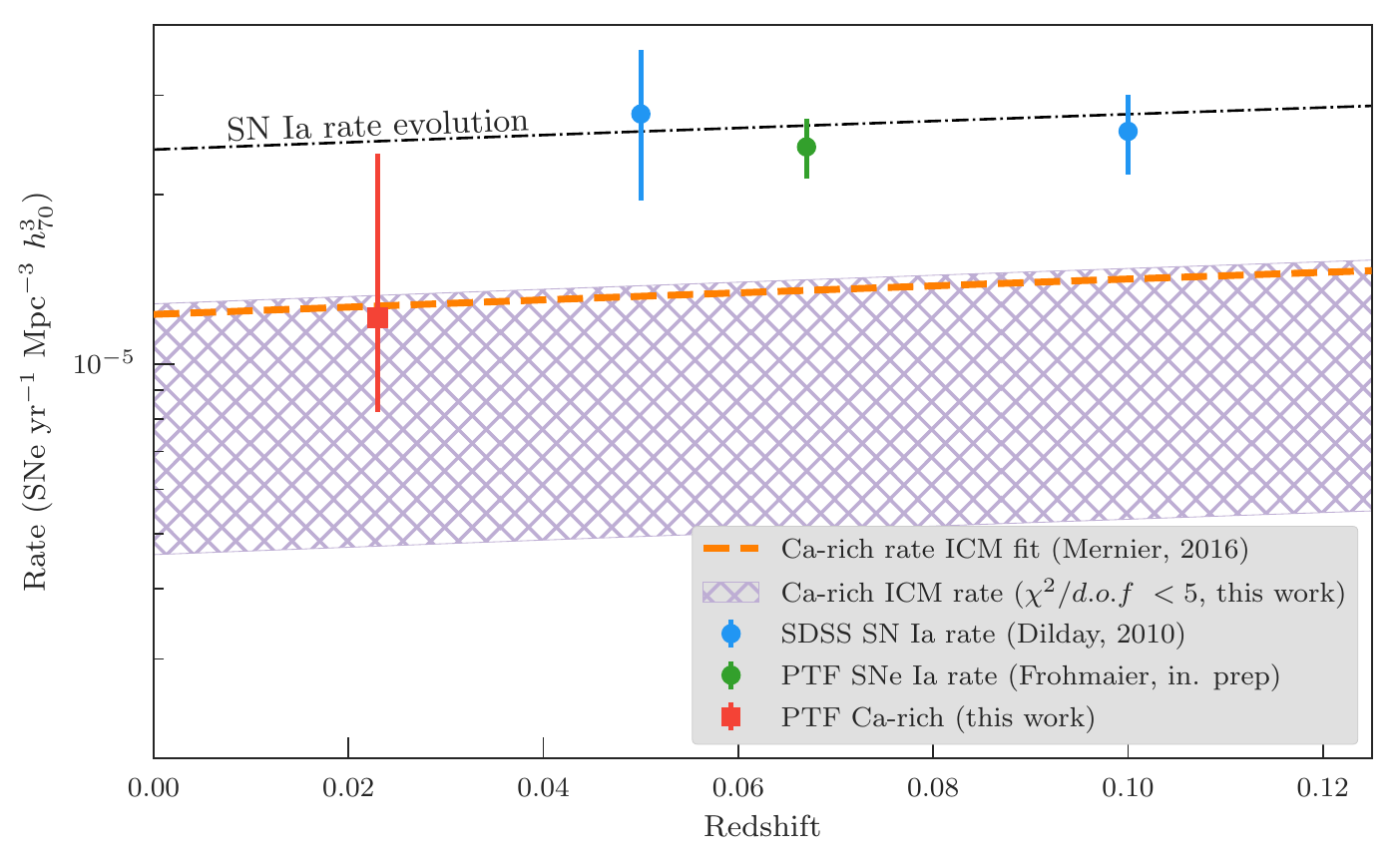}
\caption{Volumetric rate of various thermonuclear supernovae in the local universe. Our new Ca-rich measurement is shown as the red square. The circular points show SN Ia rates from SDSS-SN \citep[][blue points]{2010ApJ...713.1026D}, and the green point is from PTF (Frohmaier et al., in prep). The purple hatched region shows the possible range of Ca-rich rates from our matching of different SN class elemental yields to ICM abundance observations. The dashed orange line shows the Ca-rich rate prediction by \citet{2016A&A...595A.126M} from their best-fit to the ICM data.}
\label{fig:ca_rates}
\end{figure*}

\section{Intra-cluster Medium Calcium Abundances}
\label{sec:ca_abundances}

In this section we discuss the implications of our rate calculation with respect to the so-called `Calcium Conundrum'. Observations of the ICM show a Ca/Fe overabundance when expected elemental yields from both SNe Ia and core-collapse SNe are included in the enrichment \citep[e.g.][]{2007A&A...465..345D}. It has been shown that a non-negligible contribution from Ca-rich SNe can potentially resolve this observed discrepancy \citep{2014ApJ...780L..34M,2016A&A...595A.126M}. 

We use the latest ICM abundance measurements \citep{2016A&A...592A.157M} of nine elements X=(O, Ne, Mg, Si, S, Ar, Ca, Fe, and Ni) from XMM-Newton observations of 44 galaxy clusters. We then build on the work of \citet{2016A&A...595A.126M}, using their \texttt{abunfit} routine, to estimate the ratio of different SN classes that contribute to the ICM pollution. We do not attempt to definitively state the relative rates of the different SN types, rather we investigate whether the contribution of Ca-rich SNe is compatible with our rate result.

The asymptotic elemental mass yields for our SN Ia population are described by the Chandrasekhar mass (M$_\mathrm{Ch}$) N100 models of \citet{2013MNRAS.429.1156S} and the latest sub-Chandrasekhar mass (sub-M$_\mathrm{Ch}$) models of \citet{2017arXiv170601898S}. We select our sub-M$_\mathrm{Ch}$ models to have an initial metallicity of 0.5 or 1Z$_\odot$ and C/O WD masses of 0.9 or 1.0M$_\odot$ - which best describe `normal' SNe Ia. The implementation of both the core-collapse SNe (CCSNe) \citep{2013ARA&A..51..457N} and the Ca-rich Helium shell detonation models \citep{2011ApJ...738...21W} into \texttt{abunfit} are described in \citet[][]{2016A&A...595A.126M}. The initial metallicities allowed for the CCSNe are Z$_\mathrm{int}=(0, 0.001, 0.004, 0.008, 0.02, 0.05)$, and a Salpeter initial mass function is assumed.

The \texttt{abunfit} routine performs a least squares minimization to find the relative ratios of the SN classes that contribute to the X/Fe ICM abundances. The production of Mn and Ni is especially sensitive to the initial metallicity for the SNe Ia and the progenitor model used, furthermore, Ni is poorly constrained in the ICM abundance measurements, we therefore exclude Mn and Ni from our fits. In total, 336 model combinations were fit to the ICM observations, with 149 producing physically meaningful results. Results were rejected if they required non-positive contributions from any model yield. 

We find our best-fit result ($\chi^2/d.o.f \sim 2.6$) uses a combination of the CO.5HE.2\footnote{The naming scheme for the \citet{2011ApJ...738...21W} models represent the mass of the C/O WD core and masses of additional layers. For example, a 0.5M$_\odot$ C/O WD with a 0.2M$_\odot$ Helium shell is listed as CO.5HE.2} Ca-rich model, the Z$_\odot$ CCSN model, the N100 SN Ia model, and the Z$_\odot$, 1M$_\odot$ sub-M$_\mathrm{Ch}$ SN model. Our next best-fitting ($\chi^2/d.o.f \sim 3.9$) Ca-rich model, CO.5HE.2N.02, requires the same CCSN and M$_\mathrm{Ch}$ models as before, but a different sub-M$_\mathrm{Ch}$ model (Z$_\odot$, 0.9M$_\odot$). 

\begin{figure}
\centering
\includegraphics[width=0.95\linewidth]{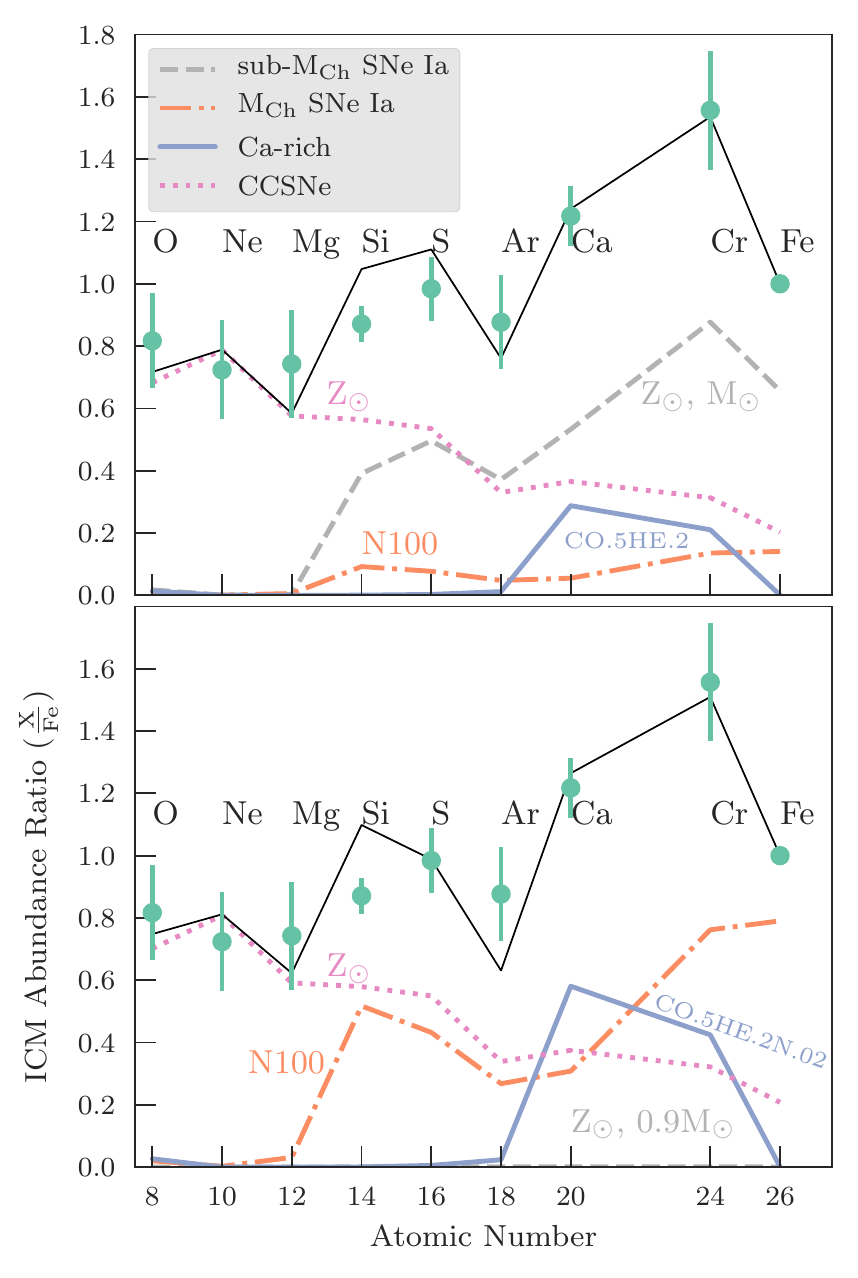}
\caption{The observed ICM abundances from \citet{2016A&A...592A.157M} are shown by the green points and the relative SN contributions by the various colored lines. The overall contribution from our SN models is shown by the solid black line. The top panel represents our best-fit combination of SNe and requires a dominant population of sub-M$_\mathrm{Ch}$ SNe Ia. The next best Ca-rich model is shown in the bottom panel and requires a negligible contribution from sub-M$_\mathrm{Ch}$ SNe Ia. We find that the required Ca-rich rate to explain these abundances is highly dependent on initial metallicity/mass of our other SN classes, but that in all cases a significant Ca-rich population is required.}
\label{fig:ICM_Abundance}
\end{figure}

The relative elemental yields for these results are shown in Figure~\ref{fig:ICM_Abundance} and clearly confirm previous claims \citep{2014ApJ...780L..34M,2016A&A...595A.126M} that the ICM measurements require a non-negligible contribution from Ca-rich SNe to explain the Ca/Fe abundance ratio.

However, we note that while a contribution from a Ca-rich model is required to produce good fits to the ICM, their relative contribution can vary significantly depending on the CCSN and SN Ia models we select. We demonstrate this by trialling different CCSN and SN Ia model combinations for a fixed Ca-rich model. We then calculate the ratio of Ca-rich SNe to SN Ia and scale the SN Ia rate evolution \citep[][Frohmaier et al. in prep.]{2010ApJ...713.1026D} to estimate the Ca-rich rate. Our results are shown in Figure~\ref{fig:ca_rates}, where any fit utilizing the CO.5HE.2 or CO.5HE.2N.02 model, with a $\chi^2/d.o.f\ < 5$ are considered. These results place the Ca-rich rate within the range of 19\%--53\% of the SN Ia rate. This is compatible with both our rate from Section~\ref{sec:rate}, and the rate prediction of \citet{2016A&A...595A.126M}. They found Ca-rich SNe made-up $\sim$34\% of their `thermonuclear' population, corresponding to a rate $\sim$51\% of the SN Ia rate.

Finally, we note that directly comparing SN rates derived from ICM abundances, and SN rates from direct observations, is a limited approach. The ICM abundance SN rates only trace events that contribute to this environment, and not elemental yields trapped in a galaxy potential. This would not necessarily be a problem if the spatial distribution of all SN classes were identical. However, as we have demonstrated in Section~\ref{sec:Eff_Environment}, Ca-rich SNe preferentially occur at large distances from their host - making them more effective polluters of the ICM. The SN rates measured using the ICM abundance method are also very dependent on the elemental yields estimated from the various explosion models.

\section{Conclusion}
\label{sec:conclusion}

We have presented three results relating to Ca-rich faint-and-fast transients; (i) using detailed simulations of PTF, we have calculated the volumetric rate of Ca-rich transients to be 33--94\% per cent of the SN Ia rate, higher than previously suggested; (ii) we confirm, using detection efficiencies from PTF, that Ca-rich events have an intrinsic preference for locations significantly offset from their host galaxy centers; (iii) we demonstrate that a non-negligible contribution from Ca-rich transients, compatible with our rate, can explain the observed ICM elemental abundances.

Attempts to unveil the intriguing nature of these transients will benefit significantly from upcoming high cadence, wide-area (ZTF) and deep (LSST) sky surveys, where a larger number of objects can be discovered. Key to this understanding will be early detections so that follow-up spectra and multi-color photometric observations can be obtained. Ultimately, studies on light curve diversity and evolution will lead to tighter constraints on the explosion models and the asymptotic elemental yields for ICM abundance analyses. Of course, large samples of Ca-rich events will also improve the statistical uncertainties associated with the rate calculations and offer better insights into the host environments of these transients.

\acknowledgments
We thank F. Mernier for making the \texttt{abunfit} code publicly available, and for the support we received in using it. We extend our gratitude to K. Shen for providing us with the decayed yields of the sub-M$_\mathrm{Ch}$ SN models. We also thank M. Kasliwal for useful comments on the manuscript prior to submission. Finally, we thank the organizers and participants of the Munich Institute for Astro- and Particle Physics (MIAPP) workshop `The Physics of Supernovae' for the stimulating discussions that started this work. M.S. acknowledges support from EU/FP7-ERC grant No. [615929]. KM acknowledges support from the STFC through an Ernest Rutherford Fellowship. We acknowledge the use of the IRIDIS High Performance Computing Facility, and associated support services at the University of Southampton, in the completion of this work. PEN acknowledges support from the DOE through DE-FOA-0001088, Analytical Modeling for Extreme-Scale Computing Environments. This research used resources of the National Energy Research Scientific Computing Center, a DOE Office of Science User Facility supported by the Office of Science of the U.S. Department of Energy under Contract No. DE-AC02-05CH11231.

We thank the anonymous referee for useful comments.

\vspace{5mm}
\facilities{PO:1.2m}

\software{astropy \citep{2013A&A...558A..33A},  
astropy/photutils \citep{larry_bradley_2016_164986},
abunfit \citep{2016A&A...595A.126M},
Matplotlib \citep{Hunter:2007},
NumPy and SciPy \citep{5725236},
LMFIT \citep{newville_2014_11813}
          }
          
\bibliographystyle{aasjournal}

\end{document}